\documentstyle[a4,11pt,epsf]{article}
\epsfverbosetrue
 0

\newcommand{\beq}{\begin{equation}}
\newcommand{\eeq}{\end{equation}}

\begin{document}

\begin{titlepage}
 
\vspace{5mm}
 
\begin{center}
{
 \huge 
       The Strength of First and Second Order Phase 
  \\[3mm]
       Transitions from Partition Function Zeroes
}
\\[15mm]
{\bf 
W. Janke$^{\rm{a}}$ 
and 
{R. Kenna$^{\rm{b}}$},
\\
$^{\rm{a}}$ 
Institut f\"ur Theoretische Physik, Universit\"at Leipzig,
\\Augustusplatz 10/11, 04109 Leipzig, Germany\\
$^{\rm{b}}$ 
School of Mathematics, Trinity College Dublin, Ireland
} 
\\[3mm]~\\ 
July 2000
\end{center}
\begin{abstract}
We present a  numerical technique employing the density
of partition function zeroes (i) to distinguish between phase 
transitions
of first and higher order, (ii) to examine the crossover between such
phase transitions and (iii) to measure the strength 
of first and second order phase transitions in the form of latent heat
and critical exponents. 
These techniques are demonstrated in applications to a number of models
for which zeroes are available.
\end{abstract}


%
\end{titlepage}

\newpage

\section{Introduction}
\label{introduction}
\setcounter{equation}{0}

A long standing concern of statistical physics
 is how best to 
distinguish between phase transitions of first and second order 
(for recent reviews, see \cite{Bi92,Ja94,Bi95}). 
First order phase transitions involve the coexistence of two
distinct phases and are characterised 
in the infinite-volume limit
by the existence of a discontinuity in the first derivative of 
the free energy.  For temperature (energy) driven first order 
phase transitions this discontinuity in the internal energy is 
the latent heat while for field (order parameter)
driven first order phase transitions the discontinuity is the 
spontaneous magnetization. 
The specific heat and the magnetic susceptibility -- the second 
derivatives of the free energy with respect to the temperature and 
external field respectively -- then exhibit delta 
function singularities in these cases. 
This  contrasts with the case of a phase transition of second order, 
where the appropriate second derivative of free energy diverges
along with the correlation length. This divergence is characterised by
critical exponents labelling the universality class of the system.
The infinite volume system does not anticipate the onset of a phase 
transition of first order as the transition region is approached. 
There is therefore no critical region and no critical exponents in 
the usual sense 
of second and higher order phase transitions.

Finite volume systems do however anticipate the presence of the phase
transition as the thermodynamic limit is approached for both of the 
above scenarios.
Numerical methods of identifying the order of a phase transition
exploit this anticipation by including finite size scaling (FSS)
of extrema of thermodynamic quantities  which are singular
in the thermodynamic limit at the transition point. The counterparts of 
these singularities in  finite systems are smooth peaks, 
the shapes of which depend on the order and the strength of
the phase transition.
The theory and use of FSS is by now well established for
both first \cite{CLB,Borgs,LeKo91,BoJa92,Ja93} and second order 
\cite{Fisher,Barber,Privman} scenarios.

An increasingly popular alternative to the analysis of these extrema
is the use of the FSS behaviour of partition function zeroes 
\cite{LY,Fi64,IPZ,GlPr86,KeLa93}. 
For field driven phase transitions one is interested in the 
Lee--Yang zeroes in the plane of complex external magnetic field \cite{LY}.
For temperature driven phase transitions the Fisher zeroes in the complex 
temperature plane are relevant \cite{Fi64}.
In this case, FSS of the first Fisher zero
yields the correlation length critical exponent $\nu$. By formally
identifying $\nu$ with $1/d$ in the first order case, FSS of
this first zero can often be used to discriminate between the two
types of phase transition. The strength of the transition, on the
other hand,  has to 
our knowledge heretofore not been determined from properties of
the zeroes, but only by direct measurement 
of the latent heat or the interface tension.
Analogous statements apply in the field driven case.

Here we wish to present an alternative method to investigate the strength
of phase transitions.
This method involves the {\em{density}} of zeroes -- 
a quantity directly proportional to the latent heat
in the first order case. 
The emphasis in this paper is on interpretation rather than new 
simulations
and our motivation comes from recent experience with
numerical studies of the density of zeroes \cite{KeLa94,XY} coupled to 
the 
ongoing discussions regarding first order phase transitions 
\cite{Ja94}. The rest of the paper is organised as follows. The 
properties of
partition function zeroes are discussed in Sec.~\ref{zeroes}.
In Sec.~\ref{an} we present the results of our analysis applied to
previously published data, and Sec.~\ref{conclusions} contains our 
conclusions.

\section{Partition Function Zeroes}
\label{zeroes}

The study of partition function zeroes was pioneered by Lee and Yang
\cite{LY} in the complex external ordering variable for field driven
phase transitions.
Study of zeroes in the complex temperature plane
was initiated by Fisher \cite{Fi64}. For convenience, one
often refers to the zeroes in the complex
external field strength plane as Lee-Yang zeroes
and those in the complex temperature plane (in the 
absence of an external field) as Fisher zeroes.
The emphasis in this paper is on Fisher zeroes, although the
techniques can also be applied to Lee-Yang zeroes.

Itzykson, Pearson and Zuber \cite{IPZ} gave the FSS
of partition function zeroes for $d$-dimensional systems below the 
upper critical dimension.
They suggested a  compact description of scaling
which for the $j^{\rm{th}}$ Lee-Yang and Fisher zero 
respectively reduces to (for large $j$)
\begin{equation}
 h_j(L)
 \sim
 \left( \frac{j}{L^d} \right)^{\frac{d+2-\eta}{2d}}
\quad ,
\quad \quad 
 t_j(L)
 \sim
 \left( \frac{j}{L^d} \right)^{\frac{1}{\nu d}}
\quad .
\label{compact}
\end{equation}
Here, $L$ is the system size, 
$\eta$ is the anomalous dimension, $t = T/T_c - 1$ 
is the reduced temperature which is zero in the first formula,
and $h$ denotes the external field which is zero in the second formula.
The integer index $j$ increases with distance from the critical point.
The FSS 
 at and above the critical dimension 
was determined in \cite{KeLa93} and \cite{GlPr86} respectively.
This behaviour of partition function zeroes offers a (generally 
very accurate) way to determine (i) the critical temperature and 
(ii) the critical  exponents
which describe the thermodynamic critical properties of the system.

Previous numerical studies of partition function zeroes mostly 
employed the  following
strategies:
\\
{\em{(i)}} Determination of the distribution of zeroes to study the
 complex parameter
phase diagram and to check (in a qualitative way) if the zeroes 
appear to cross the
real axis -- evidence for the existence of a phase transition
\cite{early,Pe82,Martin,Glasgow-QED-i,Glasgow-Hubbard-i,Glasgow-QCD-i,KiCr98}.
\\
{\em{(ii)}} Quantitative measurement of (finite size) scaling 
behaviour of those
zeroes with the smallest imaginary parts to determine the critical 
exponents
\cite{KeLa93,Glasgow-QED-ii,cbl1,cbl2,cbl3,Vi91,ViAl91,AlBe90,AlBe92,AlBe92QCD}
with possibly similar analyses applied separately to higher index 
zeroes
\cite{Ma84}
or determination of impact angle
\cite{Bh90,BhBl87,KaSh88}.
\\
{\em{(iii)}} Fixed volume plots for Lee--Yang zeroes
indicating onset of non--zero density of zeroes at 
first order 
phase transitions
\cite{Glasgow-QCD-i,AbBa96}.

Despite an early study \cite{SuKa70}, it was long considered 
difficult if
not impossible to estimate the density of zeroes from their
distribution for a finite lattice \cite{Martin}.
In recent years, however, there have been some attempts to extract
the density of zeroes from numerical studies \cite{KeLa94,density}.
In view of the increasing importance attached to this approach, we
wish to suggest an appropriate way this should be done.

To this end, we
 present a novel and powerful technique with which (i) the order
of the phase transition and (ii) the strength of the phase transition can 
be determined. This means determination of the latent heat in the first 
order 
case and  measurement of the critical exponent $\alpha$ of the specific 
heat 
in the second order case. The method
also elucidates crossover between second and first order transitions.
We study the cumulative distribution of zeroes -- curves involving $L$ 
and $j$ 
collapse --
from which the density of zeroes and hence the order and strength
of the phase transition can be determined.

\subsection{Density of Zeroes}

We start with the factorised representation of the partition function, 
\begin{equation}
 Z_L(z) = A(z) \prod_{j}{\left(z-z_j(L)\right)}
\quad ,
\end{equation}
where $z$ stands generically for 
an appropriate function of temperature (in the Fisher case)
or field (in the Lee-Yang case), $L$ is the linear extent of the lattice
and $A(z)$ is a smooth function that never vanishes. 
The free energy density follows as
\begin{equation}
 f_L(z) = \frac{1}{L^d}\ln{Z_L(z)}
        = 
\frac{1}{L^d}\ln{A(z)} + \frac{1}{L^d}\sum_{j}{\ln{(z-z_j(L))}}
\label{reee}
\quad .
\end{equation}
The first term on the right contributes only to the regular part of 
thermodynamic functions 
and we henceforth drop it. The remainder, which we refer
to as $f_L^{\rm{s}}(z)$, gives rise to singular behaviour.

Following Abe \cite{Abe}, we assume the zeroes, $z_j$,
(or at least those close
to the real axis and hence determining critical behaviour) are
on a singular line for large enough $L$, 
impacting on to the real axis at an angle $\varphi$
at the critical point $z=z_c$. The singular line is parameterised by
$z = z_c + r \exp{(i \varphi)}$. 
Define the density of zeroes as
(with $z_j=z_c+r_j \exp{(i \varphi)}$)
\begin{equation}
 g_L(r) = L^{-d} \sum_{j} \delta(r - r_j(L))
\quad . 
\label{J}
\end{equation}
The free energy is
\begin{equation}
 f_L^{\rm{s}}(z) = \int_{0}^{R}{g_L(r)\ln{(z-z_c - re^{i \varphi})} dr} 
 + {\rm{c.c.}}
\quad , 
\label{R}
\end{equation}
where c.c. means complex conjugate and $R$ is some appropriate cutoff.
The cumulative distribution function of zeroes is defined as
\begin{equation}
 G_L(r)
 =
 \int_0^r{ g_L(s) d s}
\quad.
\end{equation}
For a finite size system, therefore, this is a step function with
\begin{equation}
 G_L(r) = \frac{j}{L^d}
 \quad \quad \quad {\rm{if}} \quad 
 r \in (r_j,r_{j+1})
\quad .
\label{okG}
\end{equation}
It is natural  to assume 
that at a zero, this cumulative density 
is given by the average  \cite{LY,GlPr86,2j-1}
\begin{equation}
 G_L(r_j) = \frac{2j-1}{2L^d}
\quad .
\label{III}
\end{equation}

In the thermodynamic limit and for a phase transition of 
first order Lee and Yang \cite{LY}
already showed that the density of zeroes has to be non--zero
crossing the real axis,
\begin{equation}
 g_\infty(r) = g_\infty(0) + a r^w + \dots
\quad .
\label{cros}
\end{equation}
This corresponds to the cumulative distribution of zeroes
\begin{equation}
 G_\infty(r) = g_\infty(0) r 
+ br^{w+1} + \dots
\quad ,
\label{1st}
\end{equation}  
where the slope at the origin is related to the latent heat
(or magnetization)  via \cite{LY}
\begin{equation}
  \Delta e \propto g_\infty(0)
\quad .
\label{1stl}
\end{equation}
Abe \cite{Abe} has shown that the necessary and sufficient condition
for the specific heat at a second order phase transition to have the 
leading critical behaviour $C \sim t^{-\alpha}$, is 
(see also \cite{SuzukiLY})
\begin{equation}
 g_\infty (r) \propto r^{1-\alpha}
\quad .
\label{g2}
\end{equation}
If $\alpha = 0$, as is the case in the $d=2$ Ising model,
(\ref{g2}) leads to a logarithmic divergence in the specific heat 
\cite{Abe}. 
The corresponding expression for the cumulative distribution (integrated 
density) is
\begin{equation}
 G_\infty(r) \propto  r^{2-\alpha} 
\quad .
\label{2nd}
\end{equation}

Thus while the scaling behaviour of the position of the first (few) 
zeroes
in the complex temperature plane can  be used to identify  
$\nu$, the density of zeroes gives the strength of
the transition. A plot of $G_L(r_j) = (2j-1)/2L^d$ against $r_j(L)$
should ({\em{i}}) go through the origin, ({\em{ii}}) display $L$-- and
$j$-- collapse and ({\em{iii}}) reveal the order and strength of the 
phase
transition by its slope near the origin.

\subsection{Finite Size Scaling}

The above density approach yields new insights into traditional
finite size scaling which emerges quite naturally from it for both
first and second order phase transitions.
Traditionally, FSS is based on a hypothesis whereby the only 
relevant length scale of the system is the correlation length.
The FSS hypothesis can, in turn, be justified on renormalization
group grounds.

Alternatively, and in the second order scenario, one may start from
(\ref{III}) and (\ref{2nd}) with fixed $j$ 
and identify $G_L$ and $G_\infty$ for large enough $L$. 
The distance of a zero
from the critical point is
\begin{equation}
 r_j(L) \sim \left(\frac{2j-1}{L^d} \right)^{1/(2-\alpha)}
=
 \left(\frac{2j-1}{L^d} \right)^{1/\nu d}
\sim L^{-\frac{1}{\nu}}
\end{equation}
from hyperscaling. This is the standard formula for FSS of Fisher
zeroes at a second order phase transition implied by \cite{IPZ}.
It also recovers the compact expression (\ref{compact}) for high
index zeroes.
From this, FSS for all thermodynamic quantities can be found 
\cite{KeLa93,Abe,SuzukiLY}. 
For the specific heat, for example, (\ref{reee})
yields 
\begin{equation}
 C_L(z) = \frac{1}{L^d}\sum_j{\frac{1}{(z-z_j)^2}}
\quad .
\end{equation}
At $z=z_c$, and assuming that pseudocritical
behaviour is dominated by the lowest few zeroes \cite{KeLa93},
one finds the FSS of specific heat is
\begin{equation}
 C_L(z_c) 
 \sim L^{-d} (z_c-z_j)^{-2} \sim L^{-d}L^{2/\nu} = L^{\alpha / \nu}
\quad.
\end{equation}
Thus the usual formulae for leading order FSS are recovered.

Also, in the first order case, (\ref{III}) and
(\ref{1st}), with fixed  $j$ give
\begin{equation}
 r_j(L) \sim L^{-d}
\quad ,
\end{equation}
and hence $C_L(z_c) \sim L^d$,
explaining the formal identification of $\nu$ with $1/d$ 
alluded to in the introduction.

\section{Analysis of Various Models}
\setcounter{equation}{0}
\label{an}

In the following examples we perform fits to the cumulative density
of zeroes allowing for first or second order behaviour,
\begin{equation}
 G(r) = a_1 r^{a_2} + a_3
\quad .
\label{gen}  
\end{equation}
where we also allow for an additional parameter $a_3$.
The criterion for a good fit, apart from good data collapse 
(in $L$ and $j$),
is that $a_3$ be compatible with zero. 
In fact, a value of $a_3$ inconsistent with zero indicates 
the absence of a phase transition, 
for, if $a_3  > 0$, the zeroes have already crossed
the real axis corresponding to the broken phase and if 
$a_3 < 0$ the zeroes remain away from the real axis which is the
situation in the symmetric phase.
A first order phase transition is then
hinted at if $a_2 \sim 1$ for small $r$ (constant slope near the 
origin). 
Then the latent heat is proportional to the measured slope $a_1$.
A value of $a_2$ larger than
(and not compatible with) $1$ is indicative of a second order phase 
transition. In this case, the strength of the transition is
given by $\alpha = 2-a_2$.
The cumulative distribution of zeroes is given by (\ref{III})
and the parameter $r$ may be taken to be the imaginary part of the 
position of the $j^{\rm{th}}$ zero. That is, in practice, we fit to
\begin{equation} 
 \frac{2j-1}{2L^d} = G_L(r_j) = a_1 ({\rm{Im}}z_j(L))^{a_2} + a_3
\quad .
\end{equation}
A straightforward error analysis is not possible because of the
semi-discrete nature of this equation. We employ a  procedure
adapted from \cite{numrec} and which requires some explanation.

We observe that (\ref{III}) arose from the natural assumption 
that the density  associated with a zero in the finite volume 
case is obtained from averaging the exact formula (\ref{okG}). 
If one relaxes this assumption, and, since the cumulative 
density is monotonic, its actual value cannot 
deviate from the mean position (\ref{III}) by more than
$\pm 1/2L^d$.
Denote the data point corresponding to the $j^{\rm{th}}$ zero 
of the size $L$ lattice by $G_j^{\rm{obs}}(L)$. 
Inspired by \cite{numrec}, initially, assign an 
error $\sigma_j(L) = \sigma_{\rm{arb}}/L^d$ to this data point,
where $\sigma_{\rm{arb}}$ is a fixed arbitrary number.
Set
\begin{equation}
 \chi_1^2
 =
 \sum_{L,j}{
            \frac{
                  (G_j^{\rm{obs}}(L)
                  -G_j^{\rm{exp}}(L)
                  )^2
                 }{
                  \sigma_j(L)^2
                 }
           }
 =
 \sum_{L,j}{
            \frac{L^{2d}}{
                          \sigma^2_{\rm{arb}}
                         }
           (G_j^{\rm{obs}}(L)
           -G_j^{\rm{exp}}(L)
           )^2
           }
\quad ,
\end{equation}
where $G_j^{\rm{exp}}(L)$ is the expected density value coming from 
the
model (\ref{gen}) and  the sum runs over the lattice sizes and 
indices used in the fit. 
One proceeds in the usual way to fit the parameters $a_i$ in 
(\ref{gen})
by minimizing $\chi_1^2$. The errors associated with $a_i$ resulting 
from this fit are denoted $d a_i^{\rm{arb}}$.

Assume, now, each data point has, in fact, an error $\sigma/L^d$. 
The corresponding chi-squared is
\begin{equation}
 \chi_2^2
 =
 \sum_{L,j}{
            \frac{L^{2d}}{
                          \sigma^2
                         }
           (G_j^{\rm{obs}}(L)
           -G_j^{\rm{exp}}(L)
           )^2
           }
 = 
 \frac{\sigma^2_{\rm{arb}}}{\sigma^2}\chi_1^2
\quad .
\end{equation}
Assuming, the model fits well, this should be close to the number of 
degrees
of freedom $N_{\rm{dof}}$. In this case, the error assigned to each
point becomes
\begin{equation}
 \sigma^2
 =
 \sigma_{\rm{arb}}^2 \chi_1^2/N_{\rm{dof}}
 =
 \chi_1^2/N_{\rm{dof}}
\quad ,
\end{equation}
if $ \sigma_{\rm{arb}}$ is chosen to be unity.
Furthermore, the true errors associates with the fit parameters
are  
\begin{equation}
da_i =   \frac{\sigma}{\sigma_{\rm{arb}}} d a_i^{\rm{arb}}
     =   \sigma  d a_i^{\rm{arb}}
\end{equation}
if $ \sigma_{\rm{arb}}=1$.
One notes, however, that as in \cite{numrec}, this approach
prohibits an independent goodness-of-fit test.

\subsection{The Potts Model}
The $q$--state Potts model is the classic testing ground for the
analytical and numerical study of first and second order  
phase transitions. 
Despite the lack of a complete solution
for general $q$, some exact results are known. In two dimensions,
the model has a second order phase transition for $q\le 4$ and a first
order phase transition for $q>4$.
In terms of the inverse temperature, $\beta = 1/k_B T $, the 
critical temperature is given by $\beta_c = \ln(1+\sqrt{q})$
and the exact values of the
latent heat and a number of other thermodynamic quantities are known
\cite{Baexact}.
In the three dimensional Potts model, it is numerically established that 
there is a second order phase transition in the Ising case ($q=2$) and
a first order phase transition for $q \ge 3$.
For a review on the Potts model, see \cite{Wu}.

In the following examples we use the zeroes published in a number of papers.
If, for the Fisher zeroes, 
the parameter $z$ used is $\exp(-\kappa \beta)$ where
$\kappa$ is a trivial scale parameter depending on the model under 
consideration,
then the latent heat (for a first order phase transition) is \cite{LY}
\begin{equation}
 \Delta e = \kappa z_c 2 \pi g(0)
\quad ,
\label{Del}
\end{equation}
where  $z_c = \exp(-\kappa \beta_c)$ is the coupling at the phase
transition point and
$g(0)$ is the slope of the cumulative density plots.

\paragraph{The $d=2, q=10$ Potts Model:}
\begin{figure}[t]
\vspace{9cm}
\includegraphics{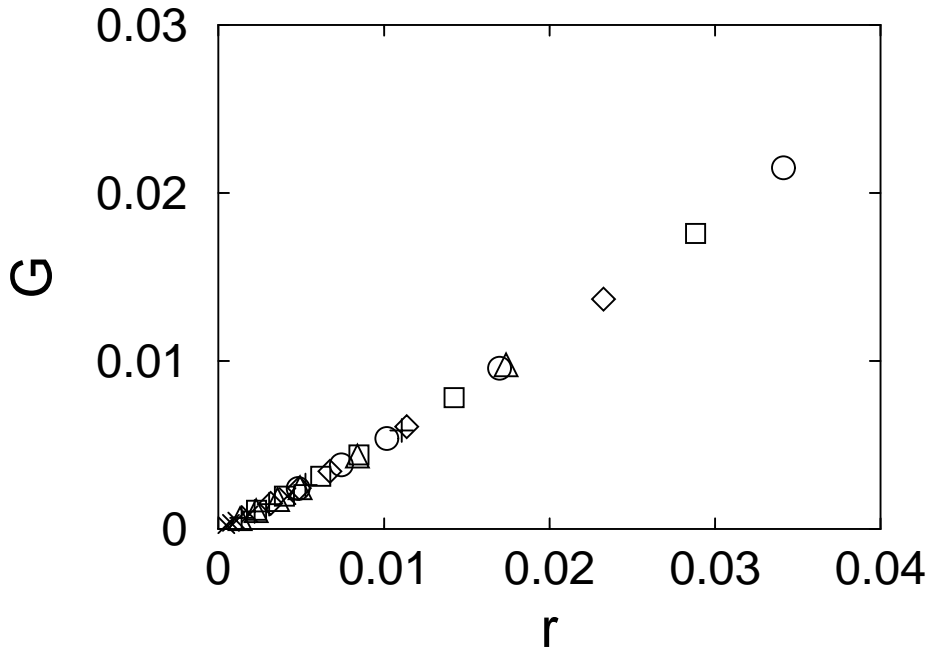}
\caption[a]{Distribution of Fisher zeroes for the 
$d=2$, $q=10$ Potts model which has a strong 
first order phase transition. The symbols 
$\times$,
$+,\bigtriangleup,\Diamond,\Box$, and 
$\put(4,4){\circle{7}}~~,$
correspond to $j=1,2,3,4,5$,  
and $6$, respectively, obtained on lattices of size $L=16$--$64$.}
\label{fig:d=2.q=10}
\end{figure}
\begin{table}[ht]
\caption{Fits of the $d=2$, $q=10$ Fisher zeroes to the cumulative
distribution $G= g(0) \times {\rm{Im}}{z}$. Here, the $N$ lowest 
zeroes from the $L=16$--$64$, $j=1$--$4$ data are used.
As the origin is approached, the resulting estimates of the
latent heat approach
the exact value 0.6961.}
\label{tab:210}
\begin{center}
\vspace{0.5cm}
\noindent\begin{tabular}{|l|llllll|}
    \hline
    \hline 
$N$       &  24      & 20       & 16       &  12      & 08       &  04
\cr
$g(0)$    & 0.501(8) & 0.486(4) & 0.479(3) & 0.471(2) & 0.469(2) & 0.463(1)   
\cr
$\Delta e$& 0.756(11)& 0.734(6) & 0.723(4) & 0.711(3) & 0.708(2) & 0.698(2)
\cr
    \hline
    \hline
\end{tabular}
\end{center}
\end{table}
Villanova \cite{Vi91} has listed the first six Fisher zeroes 
in the $z=\exp{(-\beta)}$ plane (i.e., $\kappa = 1$) for the 
two-dimensional 
$10$--state Potts model for lattice sizes $L=4,8,16,24,32,38,48$, 
and $64$ (see also \cite{ViAl91}).
In \cite{Vi91,ViAl91} a conventional FSS analysis was applied to
the first index zeroes {\em{only}}. The leading scaling behaviour 
using 
the three largest lattices yielded an acceptable fit and
$\nu^{-1} = 2.0026(11)$.
 Including the leading additive correction term gives an 
acceptable fit for $L=24,32,38,48$, and $64$ with the result   
$\nu^{-1} = 2.0028(17)$. This is good evidence for $\nu = 1/d$ and 
hence a first order phase transition. However, the determination of 
the lattice size
above which FSS (or FSS with corrections) sets in is,
by necessity,  somewhat arbitrary.
Indeed, when one extends the analysis to higher index zeroes 
(FSS applied
to each index separately), one finds that when corrections are 
ignored,
no two-parameter fit gives an acceptable result. Allowing
for corrections however (fitting to four parameters) may yield 
acceptance
indicative of a first order phase transition.
These higher index zeroes were not however analysed in 
\cite{Vi91,ViAl91}.

Our analysis of the density of zeroes begins with
Figure~\ref{fig:d=2.q=10} where all six zeroes are plotted for 
$L=16$--$64$. We do not plot the results from smaller lattices simply 
for
clarity -- including them will only give some more data points
to the right of the figure and will not affect the slope in the
region of interest, namely near the origin. Figure~\ref{fig:d=2.q=10}
displays excellent $L$-- and $j$-- collapse indicating that (\ref{III})
is the correct functional form of the density of zeroes.
Fitting (\ref{gen}) to the $L=16$--$64$, $j=1$--$4$ 
points gives 
$a_1 = 0.84(2)$, $a_2=1.10(1)$ and $a_3= 0.00004(1)$
strongly indicative of a first 
order phase transition.
Fixing $a_3=0$, and fitting for the two remaining parameters
again gives $a_2$ close to 1. Indeed, such a fit to the lowest four 
data
points gives $a_2=1.008(6)$. Fixing $a_2=1$ (and $a_3=0$)
and applying a single parameter fit to the full data set
yields $g(0) = a_1 = 0.501(8)$.
Further fits close to the origin 
yield the slopes indicated in table~\ref{tab:210}. 
From (\ref{Del}), the latent heat is 
$\Delta e = 2 \pi g(0) /(1 + \sqrt{10})$ and these 
corresponding values
 are also listed in the table and appear to converge 
to the exact value $0.6961$.

\paragraph{The $d=3$, $q=3$ Potts Model:}

\begin{figure}[t]
\vspace{9cm}
\includegraphics{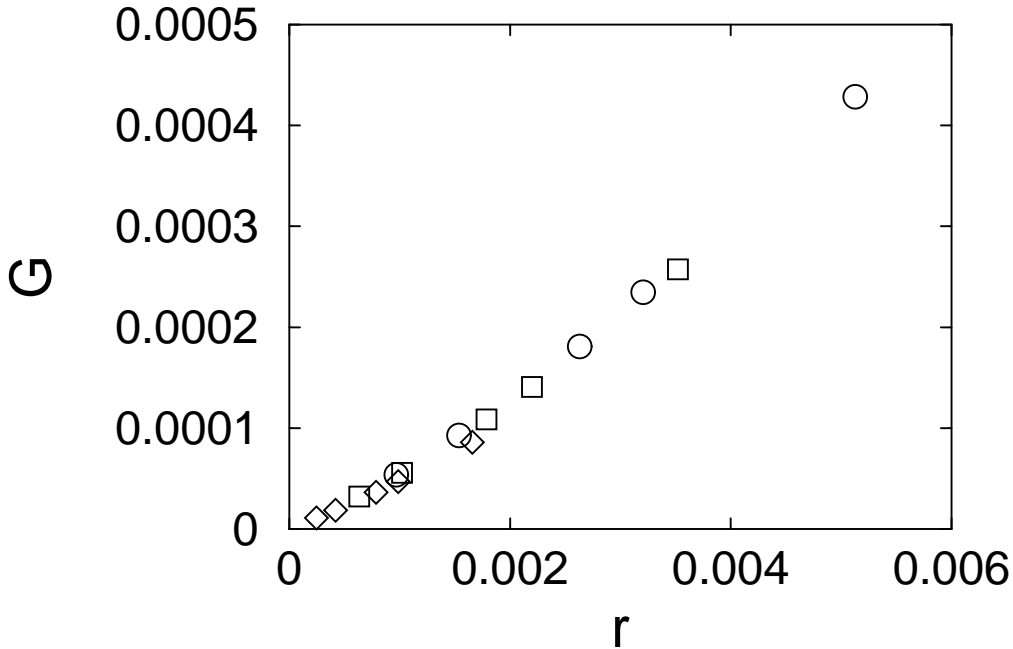}
\caption[a]{Distribution of the $L=18$--$36$ Fisher zeroes for the 
$d=3$, $q=3$ Potts model which has a weak
first order phase transition. The symbols
$\Diamond,\Box$, and  $\put(4,4){\circle{7}}~~$
correspond to $j=1,2$,  
and $3$, respectively.}
\label{fig:d=3.q=03}
\end{figure}
It is generally accepted that the three dimensional three state
Potts model has a first
order phase transition, albeit a very weak one \cite{JaVi97}. 
This is therefore
a typical model
used to test new methods to discriminate between phase transitions
of first and second order.
A list of the first few Fisher zeroes (in the $z=\exp{(-3\beta /2)}$ 
plane) for the $d=3$, $q=3$ Potts model 
for $L=10$--$36$
can be found in \cite{Vi91,ViAl91}.
In fact there is some ambiguity regarding the labelling of
the higher zeroes in \cite{Vi91,ViAl91}
as can be seen by comparing their distributions for different $L$.
To be cautious, therefore, our analysis only employs the zeroes with 
index
$j=1$--$3$.
Again in \cite{Vi91,ViAl91}, 
$\nu$ is determined by applying FSS to the first index zeroes
only. However, even with the first index zeroes, one finds
rather bad fits for any combination of more than two
lattice sizes and these are not much improved upon by allowing for 
corrections.
The way out taken by \cite{Vi91,ViAl91} is to measure $\nu$ using 
the first index
zeroes for pairs of consecutive lattice sizes in order to convince 
oneself that
$\nu$ approaches $1/d$ as $L$ increases. Their result using only the two
largest lattices is $\nu^{-1} = 2.955(26)$. Thus the evidence presented 
points towards
a first order phase transition, with, however, the caution that
``with the presented trend from other lattices, one does not feel 
confident to exclude
accidental agreement''.

Our density analysis is presented in Figure~\ref{fig:d=3.q=03}.
A 3-parameter fit to all data yields $a_3=0.000005(2)$
and becomes even closer to zero as the fit is restricted closer
to the origin.
Clearly the slope is non--zero near the origin -- the signal of a first
order phase transition. In fact, a 2-parameter fit to 
the data corresponding to $L=22,24,30,36$, $j=1$ yields
$a_1=0.070(8)$, $a_2=  1.06(2)$. Accepting that the plot
is in fact linear near the origin, and fitting for the slope only gives
$g(0) = a_1 = 0.0454(9)$.
Using $\beta_c=0.3670$ \cite{Vi91,ViAl91,JaVi97} 
and $\Delta e = (3/2)  \exp{[-(3/2) \beta_c]} 2 \pi g(0)$,
from (\ref{Del}) with $\kappa = 3/2$,
we find that the corresponding latent heat is $0.247(5)$ comparing well
with $0.2409(8)$ from \cite{Vi91,ViAl91} and with $0.2421(5)$ from the 
more
sophisticated analysis of \cite{JaVi97}.

\paragraph{The $d=3$ Ising Model:}

The Fisher zeroes in the $z=\exp({-4\beta})$ plane for the three
dimensional Ising model with 
$L=4$, $j=1$--$7$ have been determined exactly in \cite{Pe82} 
and are listed in
\cite{BhBl87} where the zeroes for $L=5$, $j=1$--$4$
are also determined. 
We also use the zeroes for $L=7$, $j=1,2$ 
from \cite{Ma84}, those for $L=6,8,10,14$, $j=1,2,3$  
which are listed in \cite{Vi91,AlBe90} and $j=1$ zeroes 
up to $L=32$ recently obtained in  \cite{AlDe00}.
\vspace{1cm}
\begin{figure}[htb]
\vspace{9cm}
\includegraphics{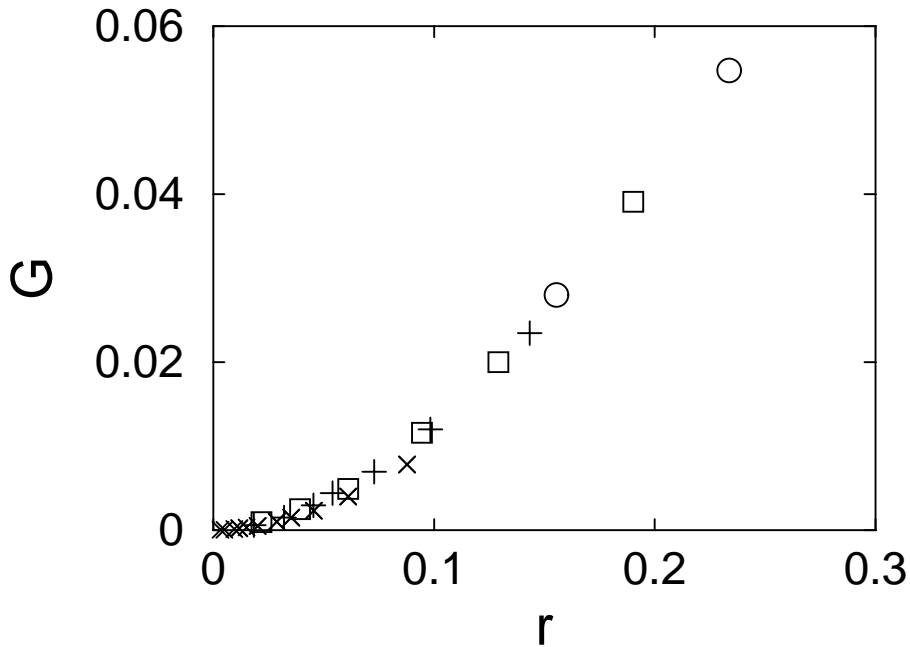}
\caption[a]{Distribution of Fisher zeroes for the 
$d=3$ Ising model which has a second
 order phase transition. The symbols 
$\times,+,\Box$, and
$\put(4,4){\circle{7}}~~~$
correspond to $j=1,2,3$,  
and $4$, respectively and the lattice size ranges from 4 to 32.}
\label{fig:d=3.q=02}
\end{figure}
The cumulative density of zeroes (using all of the above data)
is plotted in Figure~\ref{fig:d=3.q=02}.
Fitting the full set of $L=4$--$32$, $j=1$--$3$ data to  (\ref{gen})
 indicates a second 
order phase transition with $a_2=1.81(3)$, $a_3 = -0.00001(1)$.
Accepting the plot goes through the origin and applying a 2-parameter
fit to the $L\ge 6$ data gives $a_2 = 1.874(42)$ 
which corresponds to                           
 $\alpha   = 0.126(42)$.
Fitting closer to the origin (using the six data 
points corresponding to $L=10$--$32$, $j=1$ 
gives $a_2= 1.879(2)$ or                       
 $\alpha = 0.121(2)$,
close to the expected value $\approx 1/8$.
Note that the value of $\nu$ was measured to be 
$0.612(22)$, $0.6295(10)$, and $ 0.6285(19)$ in 
references \cite{Ma84}, \cite{BhBl87}, and 
\cite{Vi91,AlBe90}. The corresponding $\alpha$ 
values from hyperscaling are $\alpha = 2 -\nu d = $
                                                  $0.164(66)$, 
                                                  $0.1115(30)$, 
                                              and $0.1145(54)$,
respectively.
A weighted mean value of $\nu$ from various literature estimates
for the $d=3$ Ising model is given in \cite{JaWe00} as
$0.63005(18)$ and  corresponds to $\alpha = 0.10985(54)$.

\paragraph{The $d=2$ Ising Model:}

\begin{figure}[t]
\vspace{9cm}
\includegraphics{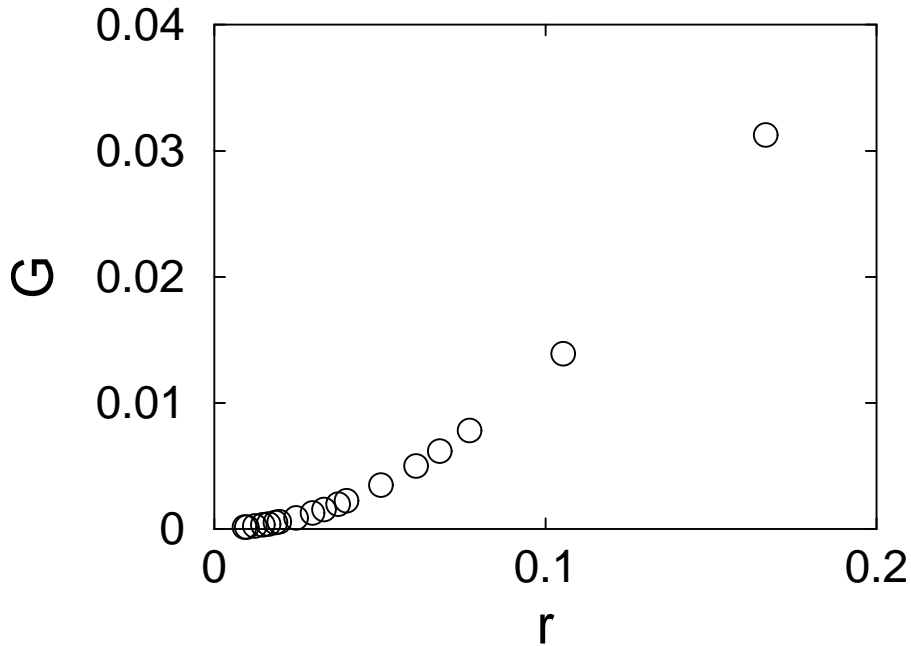}
\caption{Distribution of $j=1$ Fisher zeroes for the 
Ising model in two dimensions which has a second order 
 phase transition.
The lattice sizes range from $L=4$ to $L=64$.}
\label{fig:d=2.q=02}
\end{figure}
\begin{table}[t]
\caption{Fits of the $d=2$, $q=2$ Fisher zeroes to the cumulative
distribution $G= a_1\times ({\rm{Im}}{z})^{a_2}$. 
As the origin is approached, the parameter $a_2$ approaches 2.}
\label{tab:310}
\begin{center}
\noindent\begin{tabular}{|l|lllllllll|}
    \hline
    \hline 
$L$  &  4-64  & 10-64  & 16-64  &  20-64 & 24-64  & 32-64  &36-64 &40-64&48-64 
\cr
$a_2$&1.924(9)&1.958(4)&1.968(3)&1.972(2)&1.975(2)&1.978(1)&1.980(1)&1.981(1)&1.982(1)
\cr
    \hline
    \hline
\end{tabular}
\end{center}
\end{table}
The (numerically) exact Fisher zeroes for the Ising model in 
two dimensions for large lattices with periodic boundary conditions
(up to $L=64$) have only
recently been 
determined in  \cite{AlDeFe97}.
Their distribution 
with $z=\exp( -4 \beta)$ is plotted in Figure~\ref{fig:d=2.q=02}.

The results from two parameter fits to the $L$--ranges available
are given in table \ref{tab:310}.
Fitting 
near the origin gives $a_2 = 1.982(1)$, close to 2 which corresponds
to $\alpha = 0$. This value for $\alpha$ 
is both sufficient and necessary to attain
the logarithmic singularity of the specific heat
in the $d=2$ Ising model \cite{Abe}.

Although only the first index zeroes are available in this case,
 the
cumulative density method is still useful. 
The conventional FSS technique can only 
yield the ratio $\alpha / \nu$
from the analysis of specific heat.
Furthermore, here the specific heat leading scaling behaviour
is complicated  by logarithms
 while the density and cumulative density are 
free from such terms \cite{Abe}.
Also, FSS applied to the zeroes 
themselves (in the conventional manner) yields the correlation 
length critical exponent $\nu$, which is related to $\alpha$ in 
cases where hyperscaling holds. 
The cumulative density method is thus the only
method by which $\alpha$ can be measured  {\em{directly}} from 
finite size data.

\subsection{Lattice Gauge Theories}

Since lattice gauge theories may be viewed as statistical systems
(see, e.g., \cite{Langreview} for a review), we complete this section 
by  presenting some applications of
 our method to published zeroes of such theories.

\paragraph{The $d=3$, $SU(3)$ Model:}
     
\begin{figure}[t]
\vspace{9cm}
\includegraphics{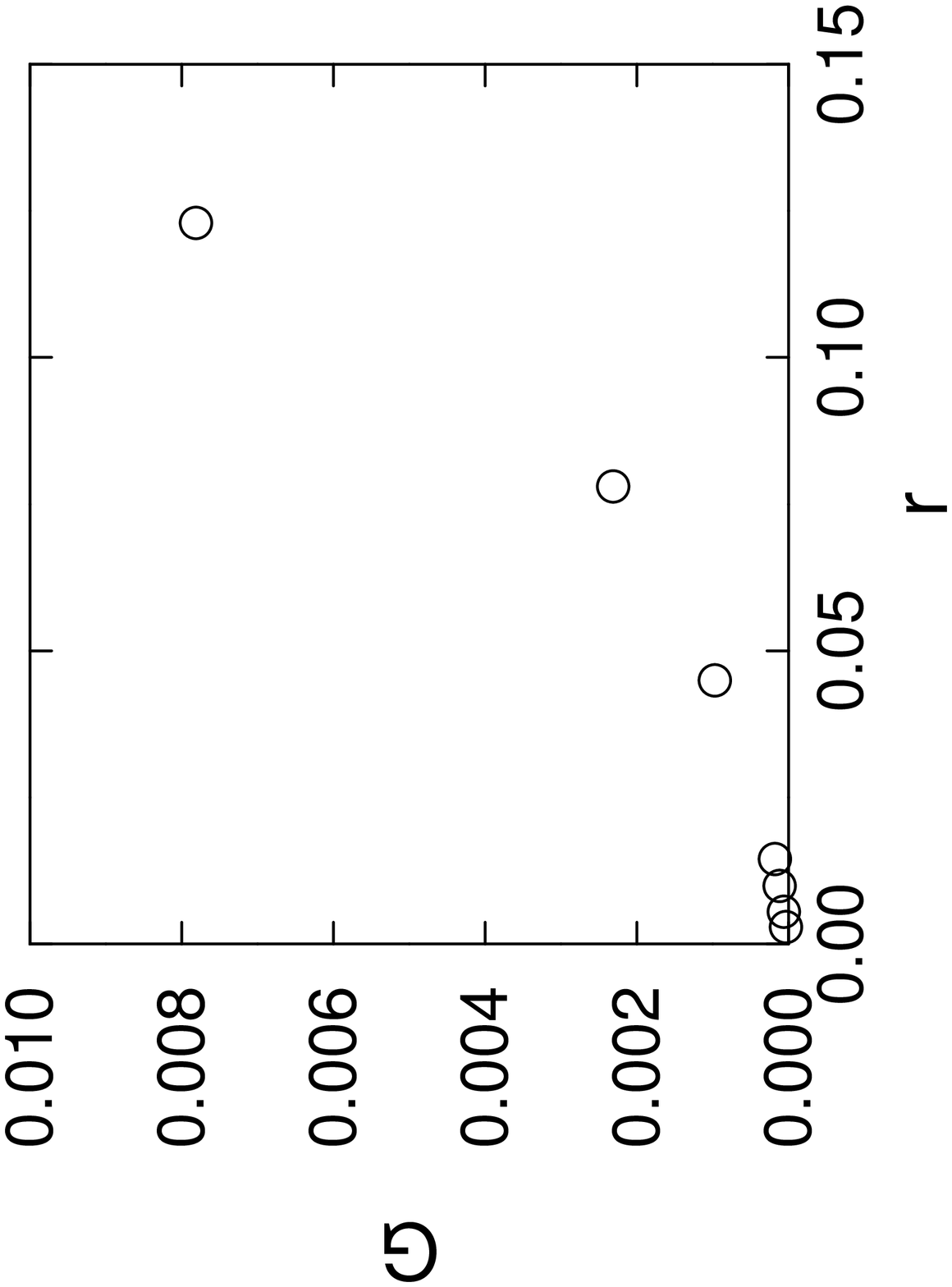}
\includegraphics{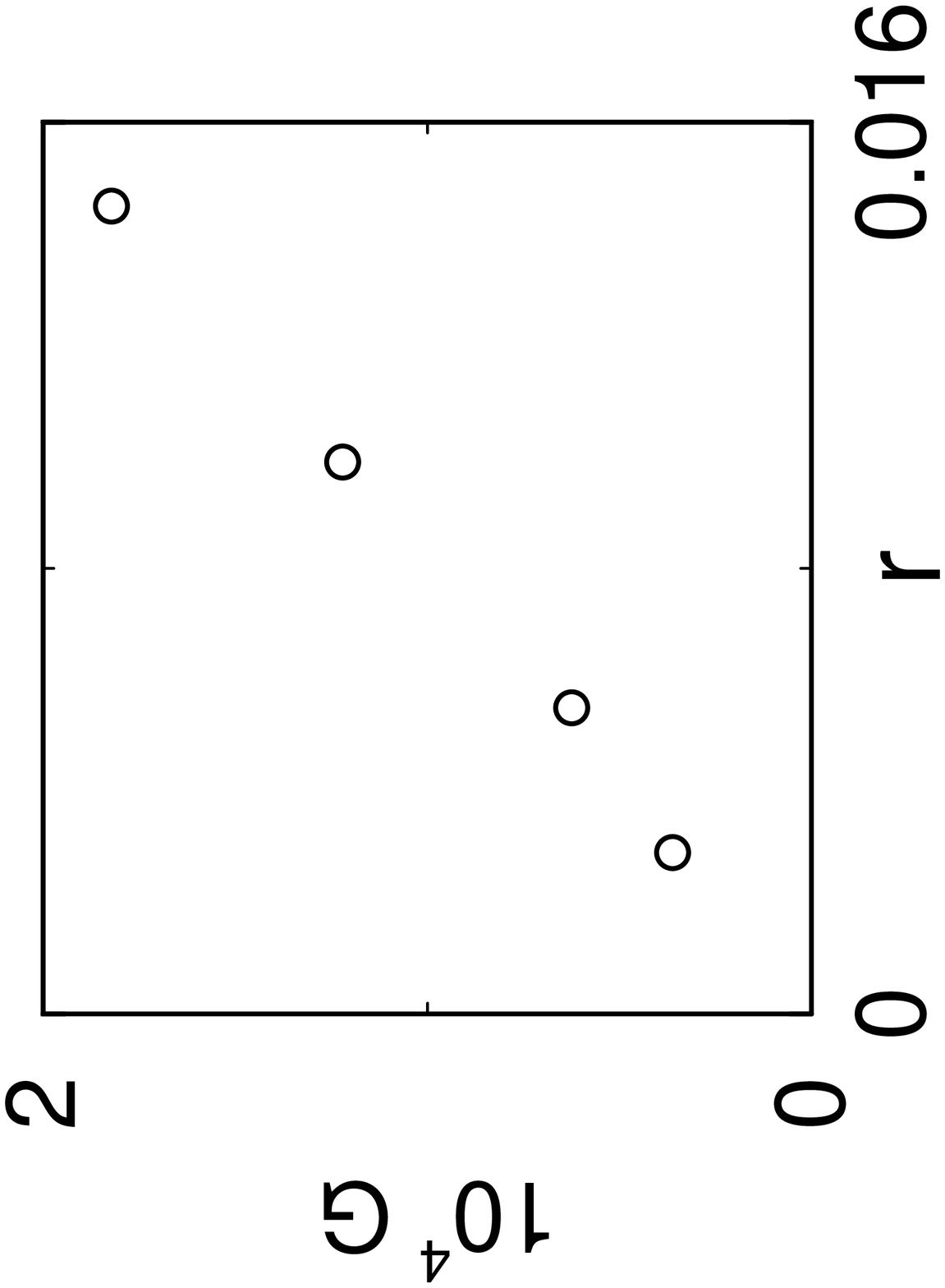}
\caption{Distribution of Fisher zeroes for the $d=3$, $SU(3)$ pure gauge
model which has a  first order 
 phase transition. The insert is a blow-up of the region near
the origin.}
\label{fig:d=3+1.su3}
\end{figure}
Alves, Berg and Sanielevici \cite{AlBe92} have analysed the $SU(3)$ 
finite temperature 
deconfining phase transition using the FSS of the lowest Fisher zeroes
for $L_tL^3$ lattices (see also \cite{KaSh88}). 
They determined the zeroes in $\beta$ 
for $L_t=4$ on lattices of size 
$L=4$, $6$, $8$, $14$, $16$,
$20$, and $24$. Their conventional FSS analysis 
gives the ``illusion of a second order phase transition''
when lattice sizes $L=4$, $6$, and  $8$
are involved in the fits.
 Claiming FSS only sets in for larger lattices, their best result
is $\nu = 0.35(2)$ for $L=14$--$24$. This value is compatible with 
$\nu = 1/d =0.33$
and thus indicative of a first order phase transition.
Figure~\ref{fig:d=3+1.su3}
is a plot of the cumulative density of zeroes 
for all lattice sizes. 
The insert  is the same 
plot using only $L=14$--$24$.
The figure, clearly supportive of a non-zero slope through the origin,
justifies restriction of the analysis to these largest lattices
and thereby elucidating the 
procedure of deciding where FSS sets in.
This slope is $0.0121(3)$,  which gives a latent heat of $0.0760(19)$. 
This is fully consistent with the
result for the latent heat
coming from conventional methods \cite{AlBe92} which is $0.0758(14)$.

\paragraph{The $d=4$, Abelian Surface Gauge Model:}
\begin{figure}[t]
\vspace{10cm}
\includegraphics{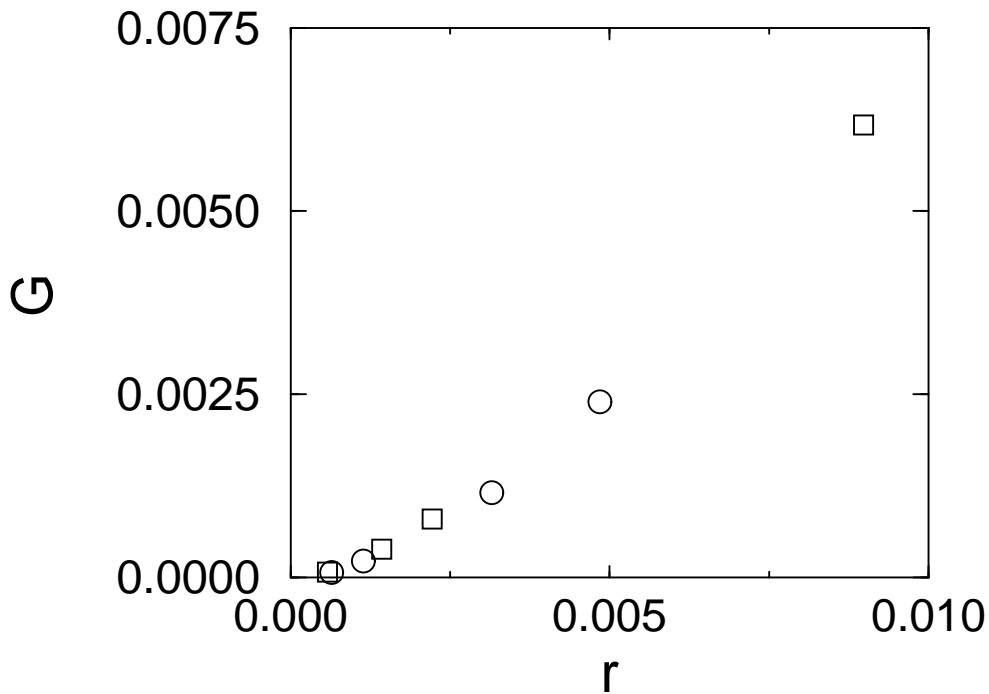}
\vspace{-1cm}
\caption[a]{Distribution of Fisher zeroes for the four dimensional abelian
surface gauge model which has a  second order (mean-field like)
 phase transition equivalent by duality to the four dimensional Ising 
model.
The $\Box$ and $\put(4,4){\circle{7}}~~~$
correspond to the $j=1$ and $j=2$ index zeroes respectively.}
\label{fig:d=4.surface.gauge}
\end{figure}
In the abelian surface gauge model, abelian gauge variables are
assigned to plaquettes of the lattice. Models such as this
are of relevance to supergravity theories amongst others.
It is known that the $d=4$ version is dual to the $d=4$ Ising
model, which, up to logarithmic corrections has mean field
critical exponents \cite{KeLa93}. One therefore expects the
$d=4$ abelian surface gauge model also to be characterised by
mean field exponents (up to corrections) with $\alpha=0, \nu =1/2$.

This model was analysed in \cite{BaVi94} where the first and
second Fisher zeroes for lattices of size $L=3$ to $L=12$ are listed. 
A conventional FSS analysis applied to the first index zero
yields the best estimate of $\nu = 0.469(17)$ 
from the two largest lattices
($L=9$ and $L=12$). It is remarked that inclusion of the smaller
lattices worsens the fit and drives $\nu$ away from $1/2$.
The appearance of a bimodal structure in the energy histograms 
was discussed in \cite{BaVi94} as a spurious indication 
of a first order transition, at odds
with the critical exponents which clearly favour second order.

A fit of the data of \cite{BaVi94} to (\ref{gen}) yields $a_2$ 
incompatible with unity
(see Figure~\ref{fig:d=4.surface.gauge}). In fact, a two parameter 
fit near the origin (using the lowest 6 zeroes) 
yields $a_2=1.90(9)$
corresponding to $\alpha = 0.10(9)$, compatible with zero.
Note, again, that only the region near the origin in 
Figure~\ref{fig:d=4.surface.gauge} is of interest.
The slope at the origin is not
compatible with a first order transition and the bimodal structure
of the energy histograms 
observed in \cite{BaVi94} can only be a (so far unexplained)
finite size effect which merits further investigation.

\section{Conclusions}
\label{conclusions}  
\setcounter{equation}{0}

The two basic questions to be addressed in the numerical study of phase
transitions are (i) identification of the order of the transition and
(ii) determination of physical quantities such as the transition point
and critical exponents or the latent heat.

We have presented a method to determine the order and strength of
phase transitions based on an analysis of
 the cumulative density of partition function zeroes. 
From the qualitative behaviour of this cumulative density 
we can distinguish between first and second order
while from the quantitative details we can extract the
latent heat in the case of first order transitions
and the specific heat  exponent $\alpha$ in the second order case.
This should be compared with traditional finite size analyses 
where only the combination $\alpha / \nu$ can be extracted
directly.

Our method meets with a high degree of success
even in the borderline cases of the $d=3$, $q=3$ Potts model
and $SU(3)$ finite temperature lattice gauge theory
where the distinction between first and second order phase
transitions is quite difficult.
There, the conclusions based on the conventional FSS analysis
are  tenuous or ambiguous.
Furthermore, the method leads to alternative insights into 
statistical physics methods in general and  especially 
illuminates the origin of finite size scaling.

\paragraph{Acknowledgements:}

R.K. would like to thank Wolfgang Grill for his hospitality during 
an extended
stay at Leipzig University in the framework of the {\em International 
Physics
Studies Program\/}.



\end{document}